\documentclass[12pt]{article}
\usepackage{amsmath}
\usepackage{bm}
\usepackage{color}

\begin{document}
\begin{center}
\begin{large}
{\bf Composite system in rotationally invariant noncommutative phase space}
\end{large}
\end{center}

\centerline {Kh. P. Gnatenko \footnote{E-Mail address: khrystyna.gnatenko@gmail.com}, V. M. Tkachuk \footnote{E-Mail address: voltkachuk@gmail.com}}
\medskip
\centerline {\small \it Ivan Franko National University of Lviv, Department for Theoretical Physics,}
\centerline {\small \it 12 Drahomanov St., Lviv, 79005, Ukraine}

\begin{abstract}
Composite system is studied in noncommutative phase space with preserved rotational symmetry.  We find conditions on the parameters of noncommutativity on which commutation relations for coordinates and momenta of the center-of-mass of composite system  reproduce noncommutative algebra for coordinates and momenta of individual particles. Also, on the conditions the coordinates and the momenta of the center-of-mass satisfy noncommutative algebra with effective parameters of noncommutativity which depend on the total mass of the system and do not depend on its composition. Besides, it is shown that on these conditions the coordinates in noncommutative space do not depend on mass and can be considered as kinematic variables, the momenta are proportional to mass as it has to be. A two-particle system with Coulomb interaction is studied and the corrections to the energy levels of the  system are found in rotationally invariant noncommutative phase space. On the basis of this result the effect of noncommutativity on the spectrum of exotic atoms is analyzed.

\end{abstract}


\section{Introduction}

In recent years studies  of noncommutative structure of space have obtained a great interest. This idea was proposed by Heisenberg and late it was presented by Snyder in his paper \cite{Snyder}. Recent interest in the noncommutativity is motivated by the development of String Theory and Quantum Gravity (see, for instance, \cite{Witten,Doplicher}).

An important problem in noncommutative space is a problem of many particles. Studies of many-particle problem give the possibility to examine the influence of noncommutativity on the properties of a wide class of physical systems.
 The problems of multiparticle quantum mechanics in a space with noncommutativity of coordinates were studied in \cite{Ho}.  In \cite{Bellucci} the system of two charged quantum particles was examined in noncommutative space. The motion of composite system in gravitational field in a space with coordinates noncommutativity was studied in \cite{GnatenkoPLA13,GnatenkoJPS13}. In paper \cite{Jabbari} the authors considered a two-body system of particles interacting through the harmonic oscillator potential on a noncommutative plane.  Also a two-particle system was considered in the context of noncommutative quantum mechanics, characterized by coordinate noncommutativity and momentum noncommutativity in \cite{Djemai}. In \cite{GnatenkoPLA17} the properties of kinetic energy of composite system in four-dimensional noncommutative phase space (2D configurational space and 2D momentum space)  were considered and the motion of the system in gravitational field was studied.
In noncommutative space-time the classical problem of many particles was examined in \cite{Daszkiewicz}. The authors considered the set of N interacting harmonic oscillators and the system of N particles moving in the gravitational field. In \cite{Daszkiewicz1} the quantum model of many particles moving in twisted N-enlarged Newton-Hooke space-time was proposed.  As an example  the system of N particles moving in and interacting by the Coulomb potential was studied.

Many-particle system was also considered in the deformed space with minimal length in \cite{Quesne,Tkachuk}. The authors found the total momenta and the coordinates of the center-of-mass position in the space and concluded that they satisfy deformed algebra with an effective parameter of deformation. On the basis of this conclusion the condition for the recovering of the equivalence principle was proposed \cite{Tkachuk}.

Canonical version of noncommutative phase space is realized with the help of the following commutation relations
  \begin{eqnarray}
[X_{i},X_{j}]=i\hbar\theta_{ij},\label{form101}\\{}
[X_{i},P_{j}]=i\hbar(\delta_{ij}+\gamma_{ij}),\label{form1001}\\{}
[P_{i},P_{j}]=i\hbar\eta_{ij}.\label{form10001}{}
\end{eqnarray}
 Here $\theta_{ij}$, $\eta_{ij}$, $\gamma_{ij}$ are elements of constant matrixes.

The noncommutative coordinates $X_i$ and noncommutative momenta $P_i$  can be represented as
\begin{eqnarray}
X_{i}=x_{i}-\frac{1}{2}\sum_{j}\theta_{ij}{p}_{j},\label{repx}\\
P_{i}=p_{i}+\frac{1}{2}\sum_{j}\eta_{ij}{x}_{j},\label{repp}
\end{eqnarray}
where coordinates $x_i$ and momenta $p_i$  satisfy the ordinary commutation relations
$[x_{i},x_{j}]=0$,  $[x_{i},p_{j}]=i\hbar\delta_{ij}$, $[p_{i},p_{j}]=0$.
Using relations (\ref{repx}), (\ref{repp}), one obtain
$[X_{i},P_{j}]=i\hbar\delta_{ij}+i\hbar\sum_k\theta_{ik}\eta_{jk}/{4}.$
Therefore, parameters $\gamma_{ij}$ are considered to be defined as
$\gamma_{ij}=\sum_k \theta_{ik}\eta_{jk}/4$ \cite{Bertolami}.

 In the canonical version of noncommutative space the rotational symmetry is not preserved \cite{Chaichian,Balachandran1}. To solve the problem of rotational symmetry breaking different noncommutative algebras were explored (see, for instance, \cite{Moreno,Galikova,Amorim,GnatenkoPLA14,GnatenkoPLA17}). Also, rotationally invariant noncommutative algebras with position-dependent noncommutativity were studied (see, for example, \cite{Lukierski,Lukierski2009,Borowiec,Borowiec1,Kupriyanov2009,Kupriyanov} and reference therein).

In our paper \cite{GnatenkoIJMPA} we studied the problem of rotational symmetry breaking in noncommutative phase space and proposed noncommutative algebra  which is rotationally invariant and equivalent to noncommutative algebra of canonical type. For this purpose we considered the idea to construct tensors of noncommutativity with the help of  additional coordinates and momenta.
In the present paper we study the problem of describing the motion of composite system in rotationally invariant noncommutative phase space proposed in \cite{GnatenkoIJMPA}.

In general case tensors of noncommutativity which correspond to different particles can be different. So, one face the problem of describing the motion of the center-of-mass of composite system in noncommutative phase space with preserved rotational symmetry.
This problem is studied in the present paper.

 In the present paper we consider a  general case when different particles feel noncommutativity with different tensors of noncommutativity. The commutation relations for coordinates and momenta of the center-of-mass of composite system and coordinates and momenta of the relative motion are analyzed in rotationally invariant noncommutative phase space. We find conditions on the parameters of noncommutativity on which coordinates and momenta of the center-of-mass satisfy noncommutative algebra with effective tensors of noncommutativity that do not depend on the systems composition. On the basis of these results a  two-particle system with Coulomb interaction is studied.

Particular case of a two-particle system with Coulomb interaction, hydrogen atom, was studied in the noncommutative space of canonical type  \cite{Ho}, in rotationally-invariant space with noncommutativity of coordinates \cite{GnatenkoU}, in noncommutative space-time \cite{Balachandran,Haghighat}. Much attention has been devoted to studies of the hydrogen atom as an one-particle problem in quantized space, see \cite{Djemai,Bertolami,Chaichian,Chaichian1,Chair,Stern,Zaim2,Adorno,Khodja,Alavi,Stern1,Moumni1,Stetsko06,Stetsko,GnatenkoPLA14,GnatenkoMPLA15,GnatenkoConf}.

In the present paper we study a system of two-particles with Coulomb interaction in rotationally-invariant space with noncommutativity of coordinates and noncommutativity of momenta.  We find corrections to the energy levels of the system and on the basis of obtained results analyze influence of noncommutativity on the spectra of exotic atoms.

The paper is organized as follows. In Section 2 we consider noncommutative phase space with preserved rotational symmetry. In Section 3 the composite system is studied. The commutation relations for coordinates and momenta of the center-of-mass, coordinates and momenta of relative motion are analyzed.
In Section 4 the hamiltonian in rotationally invariant noncommutative phase space is considered.
In Section 5 we study composite system made of two particles with Coulomb interaction and consider exotic atoms as particular cases of the system.  Conclusions are presented in Section 6.

\section{Noncommutative phase space with preserved rotational symmetry}\label{rozd2}

To preserve rotational symmetry in noncommutative phase space and construct rotationally invariant noncommutative algebra in our paper \cite{GnatenkoIJMPA}  the idea of generalization of parameters of noncommutativity $\theta_{ij}$, $\eta_{ij}$ to tensors was studied. In the paper we proposed  noncommutative algebra
  \begin{eqnarray}
[X_{i},X_{j}]=i\hbar\theta_{ij},\label{for101}\\{}
[X_{i},P_{j}]=i\hbar\left(\delta_{ij}+\sum_k\frac{\theta_{ik}\eta_{jk}}{4}\right),\label{for1001}\\{}
[P_{i},P_{j}]=i\hbar\eta_{ij},\label{for10001}{}
\end{eqnarray}
 with tensors of noncommutativity $\theta_{ij}$, $\eta_{ij}$ which are defined as
 \begin{eqnarray}
\theta_{ij}=\frac{c_{\theta} l^2_{P}}{\hbar}\sum_k\varepsilon_{ijk}\tilde{a}_{k}, \label{form130}\\
\eta_{ij}=\frac{c_{\eta}\hbar}{l^2_{P}}\sum_k\varepsilon_{ijk}\tilde{p}^b_{k}.\label{for130}
\end{eqnarray}
Here  $c_{\theta}$, $c_{\eta}$  are dimensionless constants, $l_P$ is the Planck length, $\tilde{a}_i$, $\tilde{b}_i$  $\tilde{p}^a_i$, $\tilde{p}^b_i$ are additional dimensionless coordinates and momenta conjugate of them which are governed by rotationally symmetric systems. For reason of simplicity  we consider these systems to be harmonic oscillators
 \begin{eqnarray}
 H^a_{osc}=\hbar\omega_{osc}\left(\frac{(\tilde{p}^{a})^{2}}{2}+\frac{\tilde{a}^{2}}{2}\right),\label{form104}\\
 H^b_{osc}=\hbar\omega_{osc}\left(\frac{(\tilde{p}^{b})^{2}}{2}+\frac{\tilde{b}^{2}}{2}\right),\label{for104}
 \end{eqnarray}
with
 \begin{eqnarray}
\sqrt{\frac{\hbar}{m_{osc}\omega_{osc}}}=l_{P}.\label{form200}
 \end{eqnarray}
 We also consider the frequency $\omega_{osc}$ to be very large. Therefore oscillator put into the ground state remains in it \cite{GnatenkoIJMPA}.

Operators $\tilde{a}_{i}$, $\tilde{b}_{i}$ $\tilde{p}^{a}_{i}$, $\tilde{p}^{b}_{i}$ are considered to satisfy the ordinary commutation relations $[\tilde{a}_{i},\tilde{a}_{j}]=[\tilde{b}_{i},\tilde{b}_{j}]=[\tilde{a}_{i},\tilde{b}_{j}]=[\tilde{p}^{a}_{i},\tilde{p}^{a}_{j}]=[\tilde{p}^{b}_{i},\tilde{p}^{b}_{j}]=[\tilde{p}^{a}_{i},\tilde{p}^{b}_{j}]=0,$ $[\tilde{a}_{i},\tilde{p}^{a}_{j}]=[\tilde{b}_{i},\tilde{p}^{b}_{j}]=i\delta_{ij},$
 and
$[\tilde{a}_{i},\tilde{p}^{b}_{j}]=[\tilde{b}_{i},\tilde{p}^{a}_{j}]=0.$ Also the following relations hold $[\tilde{a}_{i},X_{j}]=[\tilde{a}_{i},P_{j}]=[\tilde{p}^{b}_{i},X_{j}]=[\tilde{p}^{b}_{i},P_{j}]=0$.
So, $X_{i}$, $P_{i}$, $\theta_{ij}$, $\eta_{ij}$ satisfy the same commutation relations as in the case of the canonical version of noncommutative phase space $[\theta_{ij}, X_k]=[\theta_{ij}, P_k]=[\eta_{ij}, X_k]=[\eta_{ij}, P_k]=[\gamma_{ij}, X_k]=[\gamma_{ij}, P_k]=0.$
In this sense algebra  (\ref{for101})-(\ref{for10001}) is equivalent to algebra (\ref{form101})-(\ref{form10001}) and is rotationally invariant \cite{GnatenkoIJMPA}.

 Taking into account  expressions  (\ref{repx}), (\ref{repp}) and (\ref{form130}), (\ref{for130}) the representation  for noncommutative coordinates $X_i$ and noncommutative momenta $P_i$ reads
\begin{eqnarray}
X_{i}=x_{i}+\frac{c_{\theta} l_P^2}{2\hbar}[{\bf \tilde{a}}\times{\bf p}]_i,\label{repx0}\\
P_{i}=p_{i}-\frac{c_{\eta}\hbar}{2l_P^2}[{\bf x}\times{\bf \tilde{p}^b}]_i.\label{repp0}
\end{eqnarray}
 Coordinates $x_{i}$ and momenta $p_{i}$ satisfy the ordinary commutation relations
and commute with ${\tilde a}_{i}$, ${\tilde p}^a_{i}$, ${\tilde b}_{i}$, ${\tilde p}^b_{i}$. We would like to mention that existence of the representation guarantees that the Jacobi identity is satisfied for all possible triplets of operators.

 The algebra (\ref{for101})-(\ref{for10001}) is rotationally invariant. After rotation $X_{i}^{\prime}=U(\varphi)X_{i}U^{+}(\varphi)$, $a_{i}^{\prime}=U(\varphi)a_{i}U^{+}(\varphi)$,  $p^{b\prime}_{i}=U(\varphi)p^b_{i}U^{+}(\varphi)$  one obtains the same commutation relations
\begin{eqnarray}
[X^\prime_{i},X^\prime_{j}]=ic_{\theta} l^2_{P} \sum_k\varepsilon_{ijk}\tilde{a}^\prime_{k},\\{}
[X^\prime_{i},P^\prime_{j}]=i\hbar\left(\delta_{ij}+\frac{c_{\theta}c_{\eta}}{4}({\bf \tilde{a}^\prime}\cdot{\bf \tilde{p}^{b\prime}})\delta_{ij}-\frac{c_{\theta}c_{\eta}}{4}{\tilde a}^\prime_j{\tilde p}^{b\prime}_i\right),\\{}
[P^{\prime}_{i},P^{\prime}_{j}]= \frac{c_{\eta}\hbar^2}{l_P^2}\sum_k\varepsilon_{ijk}\tilde{p}^{b\prime}_{k}.{}
\end{eqnarray}
Here we take into account (\ref{form130}), (\ref{for130}),  the rotation operator reads $U(\varphi)=\exp(i\varphi({\bf n}\cdot{\bf L^t})/\hbar)$,
where the total angular momentum is defined as  ${\bf L^t}=[{\bf x}\times{\bf p}]+\hbar[{\bf \tilde{a}}\times{\bf {\tilde p}}^{a}]+\hbar[{\bf {\tilde b}}\times{\bf {\tilde p}}^{b}]$ \cite{GnatenkoIJMPA}.

At the end of this section it is worth mentioning that according to (\ref{repx0}) the coordinates $X_i$  depend on momenta, therefore they depend on mass. Also, noncommutative momenta (\ref{repp0})  are not proportional to mass  as it has to be. Note, that in the case when the parameters $c_{\theta}$ and $c_{\eta}$ in tensors of noncommutativity (\ref{form130}), (\ref{for130}), corresponding to a particle of mass $m$, satisfy the following conditions
\begin{eqnarray}
c_{\theta} m={\tilde \gamma}=const, \label{ccondt}\\
\frac{c_{\eta}}{m}={\tilde\alpha}=const, \label{cconde}
\end{eqnarray}
(with $\tilde \gamma$, $\tilde \alpha $ being constants which are the same for different particles), the noncommutative coordinates $X_i$ do not depend on mass and can be considered as kinematic variables in noncommutative space. Also noncommutative momenta $P_i$ are proportional to mass as it has to be.

Note, that taking into account (\ref{form130}), (\ref{for130}) conditions (\ref{ccondt}), (\ref{cconde})  are similar to the conditions $\theta m=\gamma=const,$ $\eta/m={\alpha}=const$ (here $\theta$, $\eta$ are parameters of noncommutativity corresponding to a particle of mass $m$, $\alpha$, $\gamma$ are constants which are the same for different particles) proposed in four-dimensional noncommutative phase space in \cite{GnatenkoPLA17,GnatenkoMPLA17}.  In the papers it is shown that in the case when these conditions hold  one obtain a list of important results, namely the weak equivalence principle is recovered, the properties of the kinetic energy are preserved,  the motion of the center-of-mass of composite system and the relative motion are independent and noncommutative coordinates can be considered as a kinematic variables in four-dimensional noncommutative phase space.

In the next section we  show that conditions (\ref{ccondt}), (\ref{cconde}) are important in considering a composite system in noncommutative phase space with preserved rotational symmetry.

\section{Composite system in noncommutative phase space with preserved rotational symmetry}\label{rozd4}

Let us study composite system made of $N$ particles of masses $m_n$, $n=(1...N)$ in noncommutative phase space with preserved rotational symmetry (\ref{for101})-(\ref{for10001}). We consider a general case when coordinates of different particles satisfy commutation relations (\ref{for101})-(\ref{for10001}) with different tensors of noncommutativity. It is natural to suppose that the coordinates and the momenta of different particles commute. So, let us consider the following relations
  \begin{eqnarray}
[X^{(n)}_{i},X^{(m)}_{j}]=i\hbar\delta_{mn}\theta^{(n)}_{ij},\label{ffor101}\\{}
[X^{(n)}_{i},P^{(m)}_{j}]=i\hbar\delta_{mn}\left(\delta_{ij}+\sum_k\frac{\theta^{(n)}_{ik}\eta^{(m)}_{jk}}{4}\right),\label{for1001}\\{}
[P^{(n)}_{i},P^{(m)}_{j}]=i\hbar\delta_{mn}\eta^{(n)}_{ij},\label{ffor10001}{}
\end{eqnarray}
here indexes $m,n=(1...N)$ label the particles,  and $\theta^{(n)}_{ij}$, $\eta^{(n)}_{ij}$ are the tensors of noncommutativity, corresponding to the particle of mass $m_n$.

We would like to mention here that the additional coordinates $\tilde{a}_i$ and the momenta $\tilde{p}^b_{i}$ are responsible
for the noncommutativity of the phase space. Different particles corresponds to the same noncommutative phase space. Therefore, we suppose that additional coordinates $\tilde{a}_i$ and momenta $\tilde{p}^b_{i}$ are the same for different particles. At the same time, taking into account conditions (\ref{ccondt}), (\ref{cconde}), we have that particles of different masses  feel effects of noncommutativity with different tensors. So, we consider the tensors of noncommutativity to be defined as
\begin{eqnarray}
\theta^{(n)}_{ij}=\frac{c_{\theta}^{(n)}l_P^2}{\hbar}\sum_k\varepsilon_{ijk}\tilde{a}_{k}, \label{rm13011}\\
\eta^{(n)}_{ij}=\frac{c_{\eta}^{(n)}\hbar}{l_P^2}\sum_k\varepsilon_{ijk}\tilde{p}^b_{k}, \label{r13011}
 \end{eqnarray}
where constants  $c_{\theta}^{(n)}$, $c_{\eta}^{(n)}$ correspond to a particle of mass $m_n$.

Let us consider the  total momentum ${\bf P}^c=\sum_{n}{\bf P}^{(n)}$, the
coordinates of the center-of-mass ${\bf X}^c=\sum_{n}\mu_{n}{\bf X}^{(n)}$, the momenta and  the coordinates of the relative motion ${\bf\Delta P}^{(n)}={\bf P}^{(n)}-\mu_{n}{\bf P}^c,$ ${\bf \Delta X}^{(n)}={\bf X}^{(n)}-{\bf X}^c$, defined in the traditional way. Here $\mu_n=m_n/M$, $M=\sum_{n=1}^{N}m_n$, coordinates ${\bf X}^{(n)}=(X^{(n)}_1,X^{(n)}_2,X^{(n)}_3)$ and momenta ${\bf P}^{(n)}=(P^{(n)}_1,P^{(n)}_2,P^{(n)}_3)$ satisfy (\ref{ffor101})-(\ref{ffor10001}). So, taking into account (\ref{ffor101})-(\ref{ffor10001}), we obtain the following commutation relations
\begin{eqnarray}
[X^c_i,X^c_j]=i\hbar\sum_{n}\mu_{n}^{2}\theta^{(n)}_{ij},\label{07}\\{}
[P^c_i,P^c_j]=i\hbar\sum_{n}\eta^{(n)}_{ij},\label{09}\\{}
[{X}^c_i,{P}^c_j]=i\hbar(\delta_{ij}+\sum_n\sum_k\mu_n\frac{\theta^{(n)}_{ik}\eta^{(n)}_{jk}}{4}).\label{010}{}
\end{eqnarray}
\begin{eqnarray}
[\Delta{X}_i^{(n)},\Delta{X}_j^{(m)}]=i\hbar(\delta_{nm}\theta^{(n)}_{ij}-\mu_{n}\theta^{(n)}_{ij}-\mu_{m}\theta^{(m)}_{ij}+\sum_{l}\mu_{l}^{2}\theta^{(l)}_{ij}) ,\\{}
[\Delta{P}_i^{(n)},\Delta{P}_j^{(m)}]=i\hbar(\delta_{nm}\eta^{(n)}_{ij}-\mu_m\eta^{(n)}_{ij}-\mu_n\eta^{(m)}_{ij}+\mu_n\mu_m\sum_{l}\eta^{(l)}_{ij}),\\{}
[\Delta{X}^{(n)}_i,\Delta{P}^{(m)}_j]=i\hbar\delta_{nm}(\delta_{ij}+\sum_k\frac{\theta^{(n)}_{ik}\eta^{(m)}_{jk}}{4})-i\hbar\mu_m\delta_{ij}-i\hbar\sum_k\frac{\theta^{(n)}_{ik}\eta^{(n)}_{jk}}{4}-\nonumber\\-i\hbar\sum_k\frac{\theta^{(m)}_{ik}\eta^{(m)}_{jk}}{4}+i\hbar\sum_l\sum_k\mu_l\frac{\theta^{(l)}_{ik}\eta^{(l)}_{jk}}{4},\label{08}{}
\end{eqnarray}
It is important to note that
\begin{eqnarray}
[{X}^c_{i},\Delta X_{j}^{(n)}]=i\hbar(\mu_{n}\theta^{(n)}_{ij}-\sum_m\mu_m^2{\theta}_{ij}^{(m)}),\\{}
[{P}^c_i,\Delta{P}^{n}_j]=i\hbar(\eta^{(n)}_{ij}-\mu_n\sum_{m}\eta^{(m)}_{ij}).
\end{eqnarray}
Therefore one can not consider the motion of the center-of-mass of composite system and the relative motion as independent in rotationally invariant noncommutative phase space.

Commutators for the coordinates of the center-of-mass (\ref{07}) and commutators for the total momenta (\ref{09}) equal to effective tensors of noncommutativity
\begin{eqnarray}
{\theta}^c_{ij}=\sum_n \mu^2_n\theta_{ij}^{(n)},\label{effc}\\
{\eta}^c_{ij}=\sum_{n}\eta^{(n)}_{ij},\label{eff2c}
\end{eqnarray}
which depend on the tensors of noncommutativity of a particles, forming the system $\theta_{ij}^{(n)}$, $\eta_{ij}^{(n)}$ and on their masses $m_n$. Therefore, effective tensors of noncommutativity depend on the systems composition. Commutator for coordinates and momenta of the center-of-mass equals to $i\hbar(\delta_{ij}+\sum_n\sum_k\mu_n\theta^{(n)}_{ik}\eta^{(n)}_{jk}/4)$. Note that
\begin{eqnarray}
i\hbar(\delta_{ij}+\sum_n\sum_k\mu_n\frac{\theta^{(n)}_{ik}\eta^{(n)}_{jk}}{4})\neq i\hbar(\delta_{ij}+\sum_k\frac{\theta^c_{ik}\eta^c_{jk}}{4}).
\end{eqnarray}
 So, relations (\ref{07})-(\ref{010}) can not be presented as relations of noncommutative algebra (\ref{for101})-(\ref{for10001}) with effective parameters of noncommutativity $\theta^c_{ik}$ (\ref{effc}), $\eta^c_{jk}$ (\ref{eff2c}) and do not reproduce noncommutative algebra (\ref{for101})-(\ref{for10001}).

We would like to stress that when conditions (\ref{ccondt}), (\ref{cconde}) hold, namely when
\begin{eqnarray}
c^{(n)}_{\theta}=\frac{\tilde{\gamma}}{m_n},\label{condt}\\
c^{(n)}_{\eta}=\tilde{\alpha}m_n\label{conde}
\end{eqnarray}
the tensors of noncommutativity (\ref{rm13011}), (\ref{r13011}) read
 \begin{eqnarray}
\theta^{(n)}_{ij}=\frac{\tilde{\gamma} l^2_{P}}{\hbar m_n}\sum_k\varepsilon_{ijk}\tilde{a}_{k}, \label{eform130}\\
\eta^{(n)}_{ij}=\frac{\tilde{\alpha}\hbar m_n}{l^2_{P}}\sum_k\varepsilon_{ijk}\tilde{p}^b_{k},\label{efor130}
\end{eqnarray}
and one has
 \begin{eqnarray}
 [{X}^c_{i},\Delta X_{j}^{(a)}]=[{P}^c_i,\Delta{P}^{a}_j]=0.
\end{eqnarray}
Also, in this case, the effective tensors of noncommutativity (\ref{effc}), (\ref{eff2c}) do not depend on the composition of a system. One has
\begin{eqnarray}
{\theta}^c_{ij}=\frac{\tilde{\gamma} l^2_{P}}{\hbar M}\sum_k\varepsilon_{ijk}\tilde{a}_{k},\label{eeffc}\\
{\eta}^c_{ij}=\frac{\tilde{\alpha}\hbar M}{l^2_{P}}\sum_k\varepsilon_{ijk}\tilde{p}^b_{k}.\label{eeff2c}
\end{eqnarray}
In addition, if  relations (\ref{ccondt}), (\ref{cconde}) are satisfied, taking into account (\ref{eform130}), (\ref{efor130}), from (\ref{010}) we obtain
\begin{eqnarray}
[{X}^c_i,{P}^c_j]=i\hbar(\delta_{ij}+\tilde{\gamma}\tilde{\alpha}\sum_{k,l,m}\frac{\varepsilon_{ikl}\varepsilon_{jkm}\tilde{a}_l\tilde {p}_m^b}{4})=i\hbar(\delta_{ij}+\sum_k\frac{\theta^{c}_{ik}\eta^{c}_{jk}}{4}).\label{10}{}
\end{eqnarray}

So, in the case when conditions (\ref{ccondt}), (\ref{cconde}) holds the coordinates and the momenta of the center-of-mass satisfy noncommutative algebra
\begin{eqnarray}
[X^c_i,X^c_j]=i\hbar{\theta}^c_{ij},\label{007}\\{}
[P^c_i,P^c_j]=i\hbar{\eta}^c_{ij},\label{009}\\{}
[{X}^c_i,{P}^c_j]=i\hbar(\delta_{ij}+\sum_k\frac{\theta^c_{ik}\eta^c_{jk}}{4}),\label{0010}{}
\end{eqnarray}
with effective tensors of noncommutativity (\ref{eeffc}), (\ref{eeff2c}) which depend on the total mass of the system and do not depend on its composition.

\section{Hamiltonian in rotationally invariant noncommutative phase space }\label{rozd3}

In noncommutative phase space with rotational symmetry (\ref{ffor101})-(\ref{ffor10001}) because of definition of the tensors of noncommutativity (\ref{rm13011})-(\ref{r13011}) the total hamiltonian has to be considered. This hamiltonian reads
 \begin{eqnarray}
  H=H_s+H^a_{osc}+H^b_{osc},\label{total}
 \end{eqnarray}
 here $H_{osc}^a$, $H_{osc}^b$ are given by (\ref{form104}), (\ref{for104}), $H_s$ is the hamiltonian of the system. For instance, in the case of system of particles with interaction potential energy depending on the distance between them we have
 \begin{eqnarray}
  H_s=\sum_{n}\frac{({\bf P}^{(n)})^{2}}{2m_{n}}+\frac{1}{2}\mathop{\sum_{m,n}}\limits_{m\neq n} U(|{\bf X}^{(m)}-{\bf X}^{(n)}|),
 \end{eqnarray}
 where the coordinates $X^{(n)}_{i}$ and the momenta $P^{(n)}_{i}$ satisfy noncommutative algebra  (\ref{ffor101})-(\ref{ffor10001}).

Let us rewrite hamiltonian  (\ref{total})  in the following form
  \begin{eqnarray}
H=H_0+\Delta H,\label{total1}
 \end{eqnarray}
 where
  \begin{eqnarray}
H_0=\langle H_s\rangle_{ab}+H^a_{osc}+H^b_{osc}\label{h0}\\
\Delta H= H-H_0=H_s-\langle H_s\rangle_{ab}.\label{dd}
 \end{eqnarray}
 Here we take into account that the frequency $\omega_{osc}$ of harmonic oscillators $H^a_{osc}$, $H^b_{osc}$ is large therefore oscillators put into the ground states remain in them. We use the notation $\langle...\rangle_{ab}=\langle\psi^{a}_{0,0,0}\psi^{b}_{0,0,0}|...|\psi^{a}_{0,0,0}\psi^{b}_{0,0,0}\rangle$ for averaging over the degrees of freedom of harmonic oscillators. Here $\psi^{a}_{0,0,0}$, $\psi^{b}_{0,0,0}$ being well known eigenstates of tree-dimensional harmonic oscillators $H^a_{osc}$, $H^b_{osc}$ in the ground states in the ordinary space.

 Let us find corrections caused by the term $\Delta H$ to the spectrum of  total hamiltonian (\ref{total1}).
 Note, that $[\langle H_s\rangle_{ab},H^a_{osc}+H^b_{osc}]=0$. Therefore the eigenfunctions and the eigenvalues of $H_0$ can be written as
  \begin{eqnarray}
\psi^{(0)}_{\{n_s\},\{0\},\{0\}}=\psi^s_{\{n_s\}}\psi^a_{0,0,0}\psi^b_{0,0,0},\\
E^{(0)}_{\{n_s\}}=E^s_{\{n_s\}}+3\hbar \omega_{osc}
 \end{eqnarray}
here we use notation $\psi^s_{\{n_s\}}$, $E^s_{\{n_s\}}$ for the eigenfunctions and eigenvalues of $\langle H_s\rangle_{ab}$ ($\{n_s\}$ are quantum numbers) and take into account that the oscillators are in the ground states. So, in the first order of the perturbation theory we have the following correction
   \begin{eqnarray}
\Delta E^{(1)}=\langle\psi^s_{\{n_s\}}\psi^a_{0,0,0}\psi^b_{0,0,0}|\Delta H|\psi^s_{\{n_s\}}\psi^a_{0,0,0}\psi^b_{0,0,0}\rangle=\nonumber\\=\langle\psi^s_{\{n_s\}}|\langle H_s\rangle_{ab}-\langle H_s\rangle_{ab}|\psi^s_{\{n_s\}}\rangle=0.
 \end{eqnarray}

In the second order of the perturbation theory we have
\begin{eqnarray}
\Delta
E^{(2)}=\sum_{\{n_s^{\prime}\},\{n^{a}\},\{n^{b}\}}\frac{\left|\left\langle \psi^{(0)}_{\{n_s^{\prime}\},\{n^{a}\},\{n^{b}\}}\left|
\Delta H\right|\psi^{(0)}_{\{n_s\},\{0\},\{0\}}\right\rangle\right|^{2}}{E^s_{\{n_s^{\prime}\}}-E^s_{\{n_s\}}-\hbar\omega_{osc}(n^{a}_{1}+n^{a}_{2}+n^{a}_{3}+n^{b}_{1}+n^{b}_{2}+n^{b}_{3})},\nonumber\\\label{form311}
\end{eqnarray}
here the set of numbers $\{n_s^{\prime}\}$, $\{n^{a}\}$, $\{n^{b}\}$ does not coincide with the set $\{n_s\}$,$\{0\}$, $\{0\}$, therefore for all terms in (\ref{form311}) one has $\omega_{osc}$ in the denominator.  Note, that the values $\left\langle \psi^{(0)}_{\{n_s^{\prime}\},\{n^{a}\},\{n^{b}\}}\left|\Delta H\right|\psi^{(0)}_{\{n_s\},\{0\},\{0\}}\right\rangle$ do not depend on the oscillator frequency $\omega_{osc}$ because of relation (\ref{form200}). The frequency of the oscillator is considered to be large. In the limit $\omega_{osc}\rightarrow\infty$ we have
$\lim_{\omega_{osc}\rightarrow\infty}\Delta E^{(2)}=0$.

So, up to the second order in $\Delta H$ the hamiltonian in rotationally-invariant noncommutative phase space is given by (\ref{h0}).
This conclusion will be used in the next section for studies of two-particle system with Coulomb interaction in rotationally invariant noncommutative phase space.

\section{Two-particle system with Coulomb interaction. Exotic atoms}\label{rozd4}

Let us consider a system of two particles of masses $m_1$, $m_2$ with Coulomb interaction in noncommutative phase space with preserved rotational symmetry (\ref{ffor101})-(\ref{ffor10001}).

In rotationally invariant noncommutative phase space the total hamiltonian (\ref{total}) has to be considered.
\begin{eqnarray}
 H=\frac{( {\bf P}^{(1)})^{2}}{2m_{1}}+\frac{({\bf P}^{(2)})^{2}}{2m_{2}}-\frac{\kappa}{|{\bf X}^{(1)}-{\bf X}^{(2)}|}+H^a_{osc}+H^b_{osc},\label{form777}
\end{eqnarray}
where $\kappa$ is a constant. Introducing coordinates and momenta of the center-of-mass and coordinates and momenta of the relative motion
\begin{eqnarray}
{\bf X}^c=\mu_1{\bf X}^{(1)}+\mu_2{\bf X}^{(2)},\\
{\bf P}^c={\bf P}^{(1)}+{\bf P}^{(2)},\\
{\bf X}^r={\bf\Delta X}^{(1)}-{\bf \Delta X}^{(2)}={\bf X}^{(1)}-{\bf X}^{(2)},\label{d1}\\
{\bf P}^r=\frac{1}{2}({\bf\Delta P}^{(1)}-{\bf \Delta P}^{(2)})=\mu_2 {\bf P}^{(1)}-\mu_1{\bf P}^{(2)},\label{d2}
\end{eqnarray}
 hamiltonian of the system (first three terms in (\ref{form777})) can be rewritten as follows
\begin{eqnarray}
H_s=\frac{({\bf P}^c)^{2}}{2M}+\frac{({\bf  P}^r)^{2}}{2\mu}-\frac{\kappa}{|{\bf X}^r|}.\label{form27772}
\end{eqnarray}
Here $M=m_1+m_2$, $\mu=m_1m_2/M$ and $\mu_i=m_i/M$. The operators $X^c_i$, $P^c_i$ satisfy (\ref{07})-(\ref{010}) with effective parameters of noncommutativity which in the case of two-particle system read $\theta^c_{ij}=\mu^2_1\theta^{(1)}_{ij}+\mu_2^2\theta^{(2)}_{ij}$, $\eta^c_{ij}=\eta^{(1)}_{ij}+\eta^{(2)}_{ij}$. In the case when conditions (\ref{condt}), (\ref{conde}) are satisfied, taking into account (\ref{ffor101})-(\ref{ffor10001}) (\ref{d1}), (\ref{d2}), the commutation relations for $X^r_i$, $P^r_i$  can be written as
\begin{eqnarray}
[X^r_i,X^r_j]=i\hbar \theta^r_{ij},\\{}
[P^r_i, P^r_j]=i\hbar \eta^r_{ij},\\{}
[X^r_i,P^r_j]=i\hbar(\delta_{ij}+\frac{1}{4}\sum_k \theta^r_{ik}\eta^r_{jk}).
\end{eqnarray}
with
\begin{eqnarray}
\theta^r_{ij}=\theta^{(1)}_{ij}+\theta^{(2)}_{ij},\label{tetr}\\{}
\eta^r_{ij}=\mu_2^2\eta_{ij}^{(1)}+\mu_1^2\eta_{ij}^{(2)}.{}\label{etr}
\end{eqnarray}

From  (\ref{eform130}), (\ref{efor130}), (\ref{tetr}), (\ref{etr}) we obtain that
\begin{eqnarray}
\theta^r_{ij}=\frac{c^r_{\theta} l_P^2}{\hbar}\varepsilon_{ijk}\tilde{a}_k=\frac{\tilde{\gamma}l_P^2}{\mu\hbar}\varepsilon_{ijk}\tilde{a}_k,\label{08}\\
\eta^r_{ij}=\frac{c^r_{\eta} \hbar}{l_P^2}\varepsilon_{ijk}\tilde{p}^b_k=\frac{\tilde{\alpha}\mu\hbar}{l_P^2}\varepsilon_{ijk}\tilde{p}^b_k,\label{18}
\end{eqnarray}
here $c^r_{\theta}=c^{(1)}_{\theta}+c^{(2)}_{\theta}$, $c^r_{\eta}=\mu_2^2 c^{(1)}_{\eta}+\mu_1^2 c^{(2)}_{\eta}$.
Also we can write
\begin{eqnarray}
\theta^c_{ij}=\frac{c^c_{\theta} l_P^2}{\hbar}\varepsilon_{ijk}\tilde{a}_k=\frac{\tilde{\gamma}l_P^2}{M\hbar}\varepsilon_{ijk}\tilde{a}_k,\label{28}\\
\eta^c_{ij}=\frac{c^c_{\eta} \hbar}{l_P^2}\varepsilon_{ijk}\tilde{p}^b_k=\frac{\tilde{\alpha} M\hbar}{l_P^2}\varepsilon_{ijk}\tilde{p}^b_k,\label{38}
\end{eqnarray}
 where $c^c_{\theta}=\mu_1^2c^{(1)}_{\theta}+\mu_2^2c^{(2)}_{\theta}$, $c^c_{\eta}=c^{(1)}_{\eta}+c^{(2)}_{\eta}$. Note, that from (\ref{08})-(\ref{38}) we have that in the case when conditions (\ref{condt}), (\ref{conde}) hold,  the tensors of noncommutativity corresponding to the motion of the center-of-mass $\theta^c_{ij}$, $\eta^c_{ij}$ and the tensors of noncommutativity corresponding to the relative motion $\theta^r_{ij}$, $\eta^r_{ij}$ depend on the total and reduced masses, respectively. It is important to mention that relations (\ref{ccondt}), (\ref{cconde}) are also  satisfied for  constants $c^c_{\theta}$, $c^c_{\eta}$, $c^r_{\theta}$, $c^r_{\eta}$. We have

 \begin{eqnarray}
c^c_{\theta} M=c^r_{\theta} \mu=c^{(1)}_{\theta} m_1=c^{(2)}_{\theta} m_2={\tilde \gamma}=const,\label{ccon}\\
\frac{c^c_{\eta}}{M}=\frac{c^r_{\eta}}{\mu}=\frac{c^{(1)}_{\eta}}{m_1}=\frac{c^{(2)}_{\eta}}{m_2}={\tilde\alpha}=const. \label{cco}
\end{eqnarray}

The noncommutative coordinates and noncommutative momenta can be represented as
\begin{eqnarray}
X^c_{i}=x^c_{i}-\frac{1}{2}\theta_{ij} p^c_j=x^c_{i}+\frac{1}{2}[{\bm \theta}^c\times {\bf p}^c]_i,\label{rrepx0}\\
P^c_{i}=p^c_{i}+\frac{1}{2}\eta^c_{ij} x^c_j=p^c_{i}-\frac{1}{2}[{\bm \eta}^c\times {\bf x}^c]_i,\label{rrepp0}\\
X^r_i= x^r_i-\frac{1}{2}\theta^r_{ij}p^r_j=x^r_{i}+\frac{1}{2}[{\bm \theta}^r\times {\bf p}^r]_i,\label{repd}\\
P^r_i=p^r_i+\frac{1}{2}\eta^r_{ij}x^r_j=p^r_{i}-\frac{1}{2}[{\bm \eta}^r\times {\bf x}^r]_i,\label{repd1}
\end{eqnarray}
where the components of vectors ${\bm \theta}^c$, ${\bm \eta}^c$, ${\bm \theta}^r$, ${\bm \eta}^r$ read
$\theta^c_i=\sum_{jk}\varepsilon_{ijk}\theta^c_{jk}/2$, $\eta^c_i=\sum_{jk}\varepsilon_{ijk}\eta^c_{jk}/2$, $\theta^r_i=\sum_{jk}\varepsilon_{ijk}\theta^r_{jk}/2$, $\eta^r_i=\sum_{jk}\varepsilon_{ijk}\eta^r_{jk}/2$.
Coordinates $x^r_i$, $x^c_i$ and momenta $p^r_i$, $p^c_i$ satisfy the ordinary commutation relations
\begin{eqnarray}
[x^c_{i},x^c_{j}]=[p^c_{i},p^c_{j}]=[x^r_i,x^r_j]=[p^r_i, p^r_j]=0,\\{}
[x^c_{i}, x^r_{j}]=[p^c_{i},p^r_{j}]=[x^r_i, p^c_j]=[p^r_i, x^c_j]=0,\\{}
[x^c_{i},p^c_{j}]=[x^r_{i},p^r_{j}]=i\hbar\delta_{ij}.
\end{eqnarray}
Using (\ref{rrepx0})-(\ref{repd1}) the hamiltonian can be rewritten in the following form
\begin{eqnarray}
H_s=\frac{({\bf p}^c)^{2}}{2M}+ \frac{({\bf p}^r)^{2}}{2\mu}+\frac{({\bm \eta}^c\cdot\bf{L}^c)}{2M}+\frac{[{\bm \eta}^c\times\bf{x}^c]^2}{8M}+
\frac{({\bm \eta}^r\cdot\bf{L}^r)}{2\mu}+\frac{[{\bm \eta}^r\times \bf{x}^r]^2}{8\mu}-\nonumber\\-
 \frac{\kappa}{\sqrt{ (x^r)^2-({\bm \theta}^r\cdot{\bf L}^r)+\frac{1}{4}[{\bm \theta}^r\times {\bf p}^r]^2}},\label{form2777}
\end{eqnarray}
where ${\bf L}^c=[\bf{x}^c\times {\bf p}^c]$, ${\bf L}^r=[{\bf x}^r\times {\bf p}^r]$.
Let us find the corrections to the energy levels of the system up to the second order in the parameters of noncommutativity. For this purpose let us find expansion of $H_s$ over the small parameters of noncommutativity. Note that for $1/|{\bf X}^r|$  up to the second order in the parameters of noncommutativity we have
\begin{eqnarray}
\frac{1}{|{\bf X}^r|}=\frac{1}{\sqrt{ (x^r)^{2}-({\bm \theta}^r\cdot{\bf L}^r)+\frac{1}{4}[{\bm \theta}^r\times {\bf p}^r]^{2}}}=\nonumber\\=\frac{1}{x^r}+\frac{1}{2(x^r)^{3}}({\bm{\theta}}^r\cdot {\bf L}^r)+\frac{3}{8(x^r)^{5}}({\bm \theta}^r\cdot {\bf L}^r)^{2}-\nonumber\\-
\frac{1}{16}\left(\frac{1}{(x^r)^{2}}[{\bm{\theta}^r}\times {\bf p}^r]^{2}\frac{1}{x^r}+\frac{1}{x^r}[ {\bm{\theta}}^r\times {\bf p}^r]^{2}\frac{1}{(x^r)^{2}}+\frac{\hbar^{2}}{(x^r)^{7}}[{\bm{\theta}}^r\times {\bf x}^r]^{2}\right),\label{f9}
\end{eqnarray}
where the following  notation is used
\begin{eqnarray}
x^r=|{\bf x}^r|=\sqrt{\sum_i(x^r_i)^2}.
\end{eqnarray}
 The last term in (\ref{f9}) appears because operators $(x^r)^{2}$, $[{\bm \theta}^r\times  {\bf p}^r]^{2}$ under the square root do not commute. The details of calculations needed to find expansion for operator $1/|{\bf X}^r|$  can be found in our paper \cite{GnatenkoPLA14}, where the corresponding expansion was done.

 So, for the hamiltonian $H_s$ we have the following expansion
\begin{eqnarray}
H_s=\frac{({\bf p}^c)^{2}}{2M}+\frac{({\bf p}^r)^{2}}{2\mu}-\frac{\kappa}{x^r}+\frac{({\bm \eta}^c\cdot{\bf L}^c)}{2M}+\frac{[{\bm \eta}^c\times{\bf x}^c]^2}{8M}+
\frac{({\bm \eta}^r\cdot{\bf L}^r)}{2\mu}+\frac{[{\bm \eta}^r\times{\bf x}^r]^2}{8\mu}-\nonumber\\-\frac{\kappa}{2 (x^r)^{3}}({\bm{\theta}}^r\cdot {\bf L}^r)-\frac{3\kappa}{8(x^r)^{5}}({\bm{\theta}}^r\cdot {\bf L}^r)^{2}+\nonumber\\+
\frac{\kappa}{16}\left(\frac{1}{(x^r)^{2}}[{\bm{\theta}}^r\times {\bf p}^r]^{2}\frac{1}{x^r}+\frac{1}{x^r}[ {\bm{\theta}}^r\times {\bf p}^r]^{2}\frac{1}{(x^r)^{2}}+\frac{\hbar^{2}}{ (x^r)^{7}}[{\bm{\theta}}^r\times {\bf x}^r]^{2}\right).\nonumber\\\label{ex2777}
\end{eqnarray}

Averaging over the eigenfunctions of the harmonic oscillators  $\psi^{a}_{0,0,0}$, $\psi^{b}_{0,0,0}$  we obtain
\begin{eqnarray}
\langle H_s\rangle_{ab}=\frac{( {\bf p}^c)^{2}}{2M}+\frac{(x^c)^2\langle(\eta^c)^2\rangle}{12M}+\nonumber\\
+\frac{({\bf p}^r)^{2}}{2\mu}-\frac{\kappa}{x^r}+\frac{(x^r)^2 \langle(\eta^r)^2\rangle}{12\mu}-\frac{\kappa ({L^r})^{2}\langle (\theta^r)^2\rangle}{8(x^r)^{5}}+\nonumber\\+
\frac{\kappa}{24}\left(\frac{1}{(x^r)^{2}} (p^r)^{2}\frac{1}{x^r}+\frac{1}{x^r}(p^r)^{2}\frac{1}{( x^r)^{2}}+\frac{\hbar^{2}}{(x^r)^{5}}\right)\langle (\theta^r)^2\rangle.\label{over}
\end{eqnarray}
Here we use the results of the following calculations
\begin{eqnarray}
\langle\eta^c_i\eta^c_j\rangle= \frac{(\hbar c_{\eta}^c)^2}{l_P^4}\langle\psi^{b}_{0,0,0}| \tilde{p}^{b}_i\tilde{p}^{b}_j|\psi^{b}_{0,0,0}\rangle=\frac{(\hbar c_{\eta}^c)^2}{2 l_P^4}\delta_{ij}=\frac{1}{3}\langle(\eta^c)^2\rangle\delta_{ij},\label{etac2}\\
\langle\eta^r_i\eta^r_j\rangle= \frac{(\hbar c_{\eta}^r)^2}{l_P^4}\langle\psi^{b}_{0,0,0}| \tilde{p}^{b}_i\tilde{p}^{b}_j|\psi^{b}_{0,0,0}\rangle=\frac{(\hbar c_{\eta}^r)^2}{2 l_P^4}\delta_{ij}=\frac{1}{3}\langle(\eta^r)^2\rangle\delta_{ij},\label{etar2}\\
\langle\theta^r_i\theta^r_j\rangle=\frac{(c_{\theta}^r)^2l_P^4}{\hbar^2}\langle\psi^{a}_{0,0,0}| \tilde{a}_i\tilde{a}_j|\psi^{a}_{0,0,0}\rangle=\frac{(c_{\theta}^r)^2l_P^4}{2\hbar^2}\delta_{ij}=\frac{1}{3}\langle(\theta^r)^2\rangle\delta_{ij}.\label{thetar2}
\end{eqnarray}

 In the previous section we concluded that up to the second order in $\Delta H$ the hamiltonian in rotationally invariant noncommutative phase space is given by (\ref{h0}). Note that from (\ref{ex2777}), (\ref{over}) we have that $\Delta H$  is of the first order in the parameters of noncommutativity. Therefore, up to the second order in the parameters of noncommutativity we can consider hamiltonian $H_0$  (\ref{h0})
without taking into account $\Delta H$. We have
\begin{eqnarray}
H_0=\langle H_c\rangle_{ab}+\langle H_r\rangle_{ab}+H^a_{osc}+H^b_{osc},\label{th}
\end{eqnarray}
where
\begin{eqnarray}
\langle H_c\rangle_{ab}=\frac{( {\bf p}^c)^{2}}{2M}+\frac{(x^c)^2\langle(\eta^c)^2\rangle}{12M},\label{c}\\
\langle H_r\rangle_{ab}=\frac{({\bf p}^r)^{2}}{2\mu}-\frac{\kappa}{x^r}+\frac{(x^r)^2 \langle(\eta^r)^2\rangle}{12\mu}-\frac{\kappa ({L^r})^{2}\langle (\theta^r)^2\rangle}{8(x^r)^{5}}+\nonumber\\+
\frac{\kappa}{24}\left(\frac{1}{(x^r)^{2}} (p^r)^{2}\frac{1}{x^r}+\frac{1}{x^r}(p^r)^{2}\frac{1}{( x^r)^{2}}+\frac{\hbar^{2}}{(x^r)^{5}}\right)\langle (\theta^r)^2\rangle.\label{r}
\end{eqnarray}
Operators $\langle H_c\rangle_{ab}$, $\langle H_r\rangle_{ab}$ given by (\ref{c}), (\ref{r}) correspond to the motion of the center-of-mass of the system and the relative motion, respectively.

It is worth mentioning that $[\langle H_c\rangle_{ab},\langle H_r\rangle_{ab}]=[\langle H_c\rangle_{ab},H^a_{osc}+H^b_{osc}]=0$. Therefore $\langle H_c\rangle_{ab}$ can be considered independently. Note also that the operator $\langle H_c\rangle_{ab}$ is the hamiltonian of three-dimensional harmonic oscillator with mass $M$ and frequency $\sqrt{2}\langle(\eta^c)^2\rangle/\sqrt{3}M$ with well known spectrum $\hbar\sqrt{2}\langle(\eta^c)^2\rangle(n^c_1+n^c_2+n^c_3+3/2)/\sqrt{3}M$, where $n^c_1,$ $n^c_2,$ $n^c_3$ are quantum numbers. So, because of momentum noncommutativity the motion of the center-of-mass in noncommutative phase space corresponds to the motion of harmonic oscillator with small frequency $\sqrt{2}\langle(\eta^c)^2\rangle/\sqrt{3}M$ which depend on the effective parameter of momentum noncommutativity.  Similarly, as was shown in \cite{Djemai}, the spectrum of free particle in noncommutative phase space corresponds to the spectrum of harmonic oscillator.

Operator $\langle H_r \rangle_{ab}$ (\ref{r}) can be rewritten as
\begin{eqnarray}
\langle H_r \rangle_{ab}= H^{(0)}_r+V,\label{form41}\\
 H^{(0)}_r=\frac{( { p}^r)^{2}}{2\mu}-\frac{\kappa}{x^r},\label{form9999}\\
V=V^{\eta}+V^{\theta},\\
V^{\eta}=\frac{(x^r)^2 \langle(\eta^r)^2\rangle}{3\mu},\label{ve}\\
V^{\theta}=\frac{\kappa}{x^r}-\left\langle\frac{\kappa}{X^r}\right\rangle_{ab}\label{vt}
\end{eqnarray}
here $X^r=|{\bf X}^r|$,  $1/X^r$ is given by (\ref{f9}). Up to the second order in the parameters of noncommutativity we have
\begin{eqnarray}
V^{\theta}=\frac{(x^r)^2 \langle(\eta^r)^2\rangle}{12\mu}-\frac{\kappa { (L^r)}^{2}\langle(\theta^r)^2\rangle}{8(x^r)^{5}}+\nonumber\\+
\frac{\kappa}{24}\left(\frac{1}{(x^r)^{2}}(p^r)^{2}\frac{1}{x^r}+\frac{1}{x^r}  (p^r)^{2}\frac{1}{(x^r)^{2}}+\frac{\hbar^{2}}{(x^r)^{5}}\right)\langle (\theta^r)^2\rangle.\label{vv}
\end{eqnarray}
 Operator $ H^{(0)}_r$ (\ref{form9999}) corresponds to the hamiltonian of a particle of mass $\mu$ in the Coulomb fiend in the ordinary space ($\theta_{ij}=0$, $\eta_{ij}=0$) with known eigenfunctions and eigenvalues.  Let us find corrections to the eigenvalues of the hamiltonian $H^{(0)}_r$ caused by noncommutativity. Using  results for mean values presented in  \cite{Wen-Chao}, according to the perturbation theory we have
\begin{eqnarray}
\Delta E^{(\theta\eta)}_{n,l}=\langle\psi^{(0)}_{n,l,m}|V|\psi^{(0)}_{n,l,m}\rangle=\Delta E^{(\eta)}_{n,l}+\Delta E^{(\theta)}_{n,l}
\end{eqnarray}
where corrections caused by noncommutativity of momenta are
\begin{eqnarray}
\Delta E^{(\eta)}_{n,l}=\langle\psi^{(0)}_{n,l,m}|V^{\eta}|\psi^{(0)}_{n,l,m}\rangle=\frac{\kappa a^3 n^2\langle(\eta^r)^{2}\rangle}{24\hbar^2}(5n^2+1-3l(l+1)),\label{ft411}
\end{eqnarray}
and corrections caused by the coordinates noncommutativity are as follows
\begin{eqnarray}
\Delta E^{(\theta)}_{n,l}=\langle\psi^{(0)}_{n,l,m}|V^{\theta}|\psi^{(0)}_{n,l,m}\rangle=\nonumber\\=-\frac{\hbar^{2}\kappa\langle(\theta^r)^{2}\rangle}{a^{5}n^{5}}\left(\frac{1}{6l(l+1)(2l+1)}-\frac{6n^{2}-2l(l+1)}{3l(l+1)(2l+1)(2l+3)(2l-1)}\right.+\nonumber\\\left.+\frac{5n^{2}-3l(l+1)+1}{2(l+2)(2l+1)(2l+3)(l-1)(2l-1)}-\right.\nonumber\\\left.-\frac{5}{6}\frac{5n^{2}-3l(l+1)+1}{l(l+1)(l+2)(2l+1)(2l+3)(l-1)(2l-1)}\right),\nonumber\\\label{fe411}
\end{eqnarray}
with $a=\hbar^2/\mu\kappa$. Here we use $V^{\theta}$ given by (\ref{vv}).

Note, that expression $\Delta E^{(\theta)}_{n,l}$ is divergent for the energy levels with $l=0$, $l=1$. For this levels we can not use expansion over the small parameters of noncommutativity (\ref{vv}). We are interested in finding corrections to the energy levels with $l=0$ because they are measured with high precision. So, on the basis of comparison of the corrections with experimental results a strong upper bound on the values of the parameters of noncommutativity can be obtained.

Taking into account (\ref{vt}) in the first order of the perturbation theory the corrections to the $ns$ energy levels caused by coordinates noncommutativity read
 \begin{eqnarray}
\Delta E^{(\theta)}_{n,0}=
\left\langle\psi^{(0)}_{n,0,0}\psi^{a}_{0,0,0}\left|\frac{1}{{\sqrt{(x^r)^2-({\bm \theta}^r\cdot{\bf L}^r)+\frac{1}{4}[{\bm \theta}^r\times {\bf p}^r]^2}}}-\frac{\kappa}{ x^r}\right|\psi^{(0)}_{n,0,0}\psi^{a}_{0,0,0}\right\rangle=\nonumber\\=1.72\frac{\hbar\langle\theta^r\rangle\pi \kappa}{8a^{3}n^{3}}.\nonumber\\\label{ctheta}
 \label{form721}
 \end{eqnarray}
 where $\langle\theta^r\rangle$ is defined as
 \begin{eqnarray}
\langle\theta^r\rangle =\langle\psi^{a}_{0,0,0}|\sqrt{\sum_{i}(\theta^r_{i})^{2}}|\psi^{a}_{0,0,0}\rangle=\frac{2 l_{P}^2c^r_{\theta}}{\sqrt{\pi}\hbar},\label{thetar}
\end{eqnarray}
  To write (\ref{ctheta}) we use our previous results presented in \cite{GnatenkoMPLA15,GnatenkoConf} where the calculations of corresponding integrals were done.
So, from (\ref{ft411}) and (\ref{ctheta}), the corrections to the energy levels with $l=0$ are as follows
\begin{eqnarray}
\Delta E^{(\theta\eta)}_{n,0}=\frac{a^3\kappa\langle(\eta^r)^{2}\rangle}{24\hbar^2}n^2(5n^2+1)+1.72\frac{\hbar\langle\theta^r\rangle\pi \kappa}{8a^{3}n^{3}}.\label{deltas}
\end{eqnarray}
Note that according to the obtained result for corrections the effect of noncommutativity of momenta better appears for the energy levels with large quantum numbers (corrections $\Delta E^{(\theta\eta)}_{n,l}$ are proportional to $n^4$). Energy levels with small quantum numbers are more sensitive to the noncommutativity of coordinates (corrections to the energy levels with $l=0$ (\ref{ctheta}) are proportional to $1/n^3$, corrections  to the energy levels with $l>1$ (\ref{fe411}) are proportional to $1/n^5$). In addition, in the contrast to the corrections to the energy levels with $l>1$ which are proportional to  $\langle(\theta^r)^2\rangle$ corrections to the $ns$ energy levels depend on $\langle\theta^r\rangle$. Therefore energy levels with $l=0$ are more sensitive to the noncommutativity of coordinates comparing to that with $l>1$.

Let us analyze the obtained result in particular cases of hydrogen-like atoms.
Note that the corrections to the energy levels (\ref{ft411}) caused by the momentum noncommutativity are proportional to $\langle(\eta^r)^{2}\rangle a^3$. Taking into account  (\ref{18}) we have $\langle(\eta^r)^{2}\rangle a^3\sim1/\mu$.  Corrections to the levels caused by the coordinate noncommutativity are proportional to $\langle \theta^r\rangle/a^3$ ($l=0$, (\ref{ctheta})) or $\langle(\theta^r)^{2}\rangle/a^5$ ($l>1$, (\ref{fe411})). According to (\ref{08}), we have $\langle \theta^r\rangle/a^3\sim\mu^2$ and $\langle(\theta^r)^{2}\rangle/ a^5\sim\mu^3$. So,  we can conclude that the effect of coordinate noncommutativity better appears in the spectrum of atoms with large reduced masses, namely for $ns$ energy levels with small quantum numbers $n$ of the atoms. Effect of momentum noncommutativity can be better examined considering  energy levels with large quantum numbers of atoms with small reduced masses, for example energy levels of  the hydrogen atom. On the basis of analysis presented above, we can also note that the difference in influences of coordinate noncommutativity and momentum noncommutativity on the energy levels appears better in the case of atoms with large reduced masses.

For instance, in the case of muonic hydrogen (a system of proton and muon) because of the ratio $\mu_{\mu p}/\mu_{H}\simeq m_{\mu}/m_e=206.8$ (here $m_e$, $m_{\mu}$ are the mass of electron and the mass of muon, $\mu_{\mu p}$, $\mu_{H}$ are reduced mass of muonic hydrogen and hydrogen atoms, respectively) the corrections to the energy levels with $l>1$ (\ref{fe411}) caused by the noncommutativity of coordinates are $(m_{\mu}/m_e)^3=8.8\cdot10^6$ times larger than that for the hydrogen atom. At the same time the corrections to the energy levels caused by the momentum noncommutativity (\ref{ft411}) are $206.8$ times smaller. Therefore, in the case of muonic hydrogen the difference in influences of noncommutativity of coordinates and noncommutativity of momenta on the energy levels can be better examined.

To estimate the value of the parameters of noncommutativity an assumption that corrections caused by noncommutativity are smaller than the accuracy of measurement is used. Therefore to find a strong bound on the values of  parameters of noncommutativity the results of hight precision measurements are needed.  The hydrogen atom $1s-2s$ transition frequency is measured with hight precision. The  relative uncertainty of the measurement is $4.5\times10^{-15}$ \cite{Parthey}. In our previous paper \cite{GnatenkoIJMPA} we studied hydrogen atom as an one-particle problem in rotationally invariant noncommutative phase space and estimated the values of the parameters of noncommutativity on the basis of comparison of the results for the corrections to the energy of $1s-2s$ transition with accuracy of the experimental results.  Note that  the orders of upper bounds for the parameters of noncommutativity presented in \cite{GnatenkoIJMPA} are not changed including the effect of reduced mass. This is because of the ratio $m_p/m_e=1836$, therefore $\mu\simeq m_e$.

It is interesting to study particular cases of two-particle systems which consist on the particles with close masses and study influence of noncommutativity on the spectrum of the systems.
Let us consider antiprotonic helium $\bar{p}^4He^+$.  This is an exotic atom composed of an antiproton, an electron and a helium nucleus. Antiprotonic helium  has a long life time comparing to other antiprotonic atoms. As was noted in \cite{Hayano,Hayano10} the $\bar{p}^4He^+$ transition frequency can be approximately written as known expression for the hydrogen atom transition frequency but with effective nuclear charge $Z_{eff}<2$  which describes the shielding of the nuclear charge by the electron, and with taking into account  the difference of masses of hydrogen and antiprotonic helium atoms. Therefore, to estimate the orders of the values of parameters of noncommutativity on the basis of $\bar{p}^4He^+$  transition frequencies measurements  we can use our result for the corrections (\ref{ft411}), (\ref{fe411}). Note that in comparison to the hydrogen atom $\bar{p}^4He^+$ has a large reduced mass. Therefore, taking into account analysis of dependence of corrections to the energy levels on reduced mass which are presented above, we can conclude that the influence of coordinate noncommutativity on the spectrum of antiprotonic helium  appears better than in the case of hydrogen atom. So, the antiprotonic helium is an attractive candidate for examinations of effects of noncommutativity.

 The frequency of  transition $(n, l) = (36, 34)\rightarrow(34, 32)$ of $\bar{p}^4He^+$  $f = 1522107062$ MHz is measured with the total experimental error $3.5$ MHz \cite{Hori}. On the basis of assumption that the corrections to the energy levels caused by the noncommutativity are smaller than the accuracy of measurements we can write
\begin{eqnarray}
|\Delta^{(\theta)}+\Delta^{(\eta)}|\leq3.5\textrm{MHz},\label{in}
\end{eqnarray}
where $\Delta^{\theta}$, $\Delta^{\eta}$ are corrections to the transition energy caused by the coordinates noncommutativity and momentum noncommutativity
\begin{eqnarray}
 \Delta^{\theta}=\Delta E^{(\theta)}_{36,34}-\Delta E^{(\theta)}_{34,32},\\
 \Delta^{\eta}=\Delta E^{(\eta)}_{36,34}-\Delta E^{(\eta)}_{34,32},
 \end{eqnarray}
 here  $\Delta E^{(\theta)}_{n,l}$, $\Delta E^{(\eta)}_{n,l}$ are given by (\ref{ft411}), (\ref{fe411}).
Note that if inequality $|\Delta^{(\theta)}|+|\Delta^{(\eta)}|\leq3.5$MHz is satisfied, the inequality (\ref{in}) is satisfied too. So, in order to  estimate the orders of the values of parameters noncommutativity, it is sufficiently to study the following inequalities $|\Delta^{\theta}|\leq1.75$ MHz, $|\Delta^{\eta}|\leq1.75$ MHz. It is also sufficiently to put $Z=2$,  $a=m_ea_B/m_{\bar{p}}$ (here $m_{\bar{p}}$ is the mass of antiproton, $a_B$ is the Bohr radius of the hydrogen atom)  in (\ref{ft411}), (\ref{fe411}). So, we obtain
\begin{eqnarray}
\hbar\langle\theta^r\rangle\leq10^{-27}\,\textrm{m}^{2},\label{form55}\\
\hbar\sqrt{\langle(\eta^r)^{2}\rangle}\leq10^{-50}\,\textrm{kg}^{2}\textrm{m}^{2}/\textrm{s}^{2}.\label{for55}
\end{eqnarray}
Note, that in the literature more strong restrictions on the values of parameters of noncommutativity are presented. For instance, upper bounds obtained from the spectrum of gravitation quantum well \cite{Bertolami1}, from the effects of noncommutativity on the hyperfine structure of hydrogen atom in noncommutative phase space without preserved rotational symmetry \cite{Bertolami}. The reason of results (\ref{form55}), (\ref{for55}) is a not hight precision of the measurements of the spectrum of exotic atoms, in particular $\bar{\textrm{p}}^4\textrm{He}^+$ atom. Nevertheless, it is important to note that comparing to hydrogen atom antiprotonic helium is more sensitive to the noncommutativity of coordinates because the reduced mass of $\bar{\textrm{p}}^4\textrm{He}^+$ is three orders larger. Therefore, improvement of precision of measurements of  $\bar{\textrm{p}}^4\textrm{He}^+$  spectrum will give a possibility to obtain more strong restriction on the values of  parameters of noncommutativity.

\section{Conclusions}

In the paper we have considered rotationally invariant space with noncommutativity of coordinates and noncommutativity of momenta (\ref{for101})-(\ref{for10001}) which was proposed in \cite{GnatenkoIJMPA}. The rotationally invariant noncommutative algebra is constructed on the basis of the idea of generalization of parameters of noncommutativity to  tensors. The tensors are constructed with the help of additional coordinates which are governed by a rotationally symmetric system. The noncommutative algebra is rotationally invariant and equivalent to the algebra of canonical type (\ref{form101})-(\ref{form10001}).

In the general case different particles in noncommutative phase space may feel effects of noncommutativity with different tensors of noncommutativity. Therefore there is a problem of describing the motion of the center-of-mass of a composite system in rotationally invariant noncommutative phase space. In the paper we have shown that the commutation relations for  coordinates and  momenta of center-of-mass of a system, the commutation relations for coordinates and momenta of relative motion depend on the tensors of noncommutativity of particles which form the system and on their masses, therefore depend on composition of the system. Also, we have shown that the algebra for coordinates and momenta of the center-of-mass (\ref{07})-(\ref{010}) does not reproduce algebra for coordinates and momenta of individual particles (\ref{for101})-(\ref{for10001}). The situation is changed if we consider the  tensors of noncommutativity  to be dependent on mass (\ref{eform130}), (\ref{efor130}). In the case when the conditions (\ref{ccondt}), (\ref{cconde}) hold the coordinates and momenta of the center-of-mass satisfy noncommutative algebra (\ref{007})-(\ref{0010})  with effective tensors of noncommutativity which do not depend on the composition of the system but only on its total mass (\ref{eeffc}), (\ref{eeff2c}). Besides, in the case when conditions (\ref{ccondt}), (\ref{cconde}) hold the noncommutative coordinates (\ref{repx0}) do not depend on the mass and can be considered as kinematic variables,  the noncommutative momenta (\ref{repp0}) are proportional to  mass as it has to be.

It is worth noting that the idea of dependence of parameters of quantized space on mass is not new. The idea was proposed to recover fundamental principles in noncommutative space \cite{GnatenkoPLA13,GnatenkoMPLA16}, in four-dimensional (2D configurational space and 2D momentum space) noncommutative phase phase \cite{GnatenkoPLA17,GnatenkoMPLA17}, in deformed space with minimal length \cite{Tkachuk,Quesne,TkachukFond}. We would like to stress that the conditions (\ref{ccondt}), (\ref{cconde}) are similar to the conditions on the parameters of noncommutativity $\theta m=\gamma=const$, $\eta/m=\alpha=const$, which were proposed in four-dimensional noncommutative phase space in the papers \cite{GnatenkoPLA17,GnatenkoMPLA17} in order to solve the problem of violation of the properties of the kinetic energy, violation of the weak equivalence principle, to consider noncommutative coordinates as kinematic variables.

The system of two particles with Coulomb interaction was studied in rotationally invariant noncommutative phase space. We have found corrections to the spectrum of the system caused by noncommutativity. We have obtained that corrections to the energy levels caused by noncommutativity of coordinates (\ref{fe411}), (\ref{ctheta}) and  noncommutativity of momenta  (\ref{ft411}) have different dependencies on the parameters of noncommutativity, and on  the quantum numbers. Also, the corrections have different dependencies on the parameters of a system (reduced mass $\mu$, parameter of interaction $\kappa$). This fact gives the possibility to select a system which  has a good sensitivity to the particular type of noncommutativity (noncommutativity of momenta or noncommutativity of coordinates).
Analyzing  (\ref{ft411}), (\ref{fe411}), (\ref{ctheta}), we have concluded that the effect of momentum noncommutativity better appears for the $ns$ energy levels with large quantum numbers of atoms with small reduced masses, in particular hydrogen atom. The effect of coordinate noncommutativity can be better examined considering $ns$ energy levels with small quantum numbers  of atoms with large reduced masses.

An influence of noncommutativity on the energy levels of exotic atoms  has been analyzed. We have concluded that antiprotonic helium is an attractive candidate for examinations of effects of noncommutativity of coordinates.  So, improvement of the accuracy of measurements of the  spectrum of antiprotonic helium will give the possibility to find more strong restrictions on the value of parameter of noncommutativity.

\section*{Acknowledgments}
This work was supported in part by the European Commission under the project STREVCOMS PIRSES-2013-612669 and the projects $\Phi\Phi$-63Hp, $\Phi\Phi$-30$\Phi$ (No. 0116U001539) from the Ministry of Education and Science of Ukraine.

\end{document}